# Influence of structural disorder on magnetic properties and electronic structure of $YCo_2$

**Z. Śniadecki*, N. Pierunek, B. Idzikowski, B. Wasilewski, M. Werwiński,**
Institute of Molecular Physics, Polish Academy of Sciences
M. Smoluchowskiego 17, 60-179 Poznań, Poland
**U.K. Rößler**
Leibniz Institute for Solid State and Materials Research Dresden, P.O. Box 270116, D-01171 Dresden, Germany
**Yu. Ivanisenko**
Institute of Nanotechnology, Karlsruhe Institute of Technology, Hermann-von-Helmholtz-Platz 1, D-76344 Eggenstein-Leopoldshafen, Germany

*corresponding author: Zbigniew Śniadecki, e-mail: sniadecki@ifmpan.poznan.pl

**Abstract**

In this paper the changes of magnetic properties with increasing disorder in the exchange-enhanced Pauli paramagnet $YCo_2$ are discussed. The structural disorder is initially introduced by rapid quenching, while further changes on micro-/nanoscale are caused by a high pressure torsion (HPT). Values of the magnetic moment determined for the plastically deformed ribbons reach 0.10 $\mu_B$/Co atom (for a sample subjected to the deformation at a pressure of 4 GPa) and 0.25 $\mu_B$/Co (6 GPa) at 2 K. Magnetic moment arise not only from the surface of nanocrystals but also from volume. *Ab initio* calculations explained the influence of chemical disorder and different types of structural defects on the electronic structure and magnetic properties of $YCo_2$-based Laves phases. The calculated magnetic ground states are in qualitative agreement with experimental results for all considered structures with point defects.

## Introduction

The crystalline $A_{1-x}Co_x$ alloys with a non-magnetic A element (Lu, Y, Sc, Zr, Ti, Nb) reveal a variety of magnetic orderings, from paramagnetic through metamagnetic to ferromagnetic. Their magnetic characteristics depend mainly on the stoichiometry, type of the non-magnetic element and crystal structure of a specific alloy [1–4]. An interesting representative of this class of materials is the Y-Co system, whose magnetic properties change significantly with the concentration of the constituent elements. The compounds with more than about 33 at.% of Y are paramagnetic. In contrast, the Y concentration below 33 at.% (like in $YCo_3$, $Y_2Co_7$, $YCo_5$ and $Y_2Co_{17}$) induces magnetic ordering with the effective magnetic moments that increase with decreasing Y content [5, 6]. $YCo_2$ is close to the mentioned threshold and is an exchange-enhanced Pauli paramagnet [7, 8], for which a slight increase in Co concentration leads to the transition from paramagnetic to ferrimagnetic state [6]. It crystallizes in the $MgCu_2$-type structure with the lattice parameter $a$ = 7.217 Å [9]. For $YCo_2$ the magnetic phase transition may also be field-induced and the metamagnetic transition was reported to take place at 70 T and at 10 K [10]. The ordered magnetic state may be moreover induced in $YCo_2$ by doping [11] or amorphization. In the amorphous Y-Co phase diagram [5] the magnetic transition is observed at a lower Co concentration than for the crystalline Y-Co. Thus, the amorphous and nanocrystalline $YCo_2$ systems exhibit a long-range magnetic ordering even in the absence of an external field [12,13]. The ferrimagnetic state with a high Curie temperature $T_C$ is also observed in the amorphous fine $YCo_2$ particles [12]. The XMCD measurements made on the $YCo_2$ amorphous films showed that the magnetic contribution from the Y atoms was negligible [14] and the main contribution came from Co atoms.

In the studies on the magnetic properties of the doped $YCo_2$ usually the Y atoms are substituted [15]. However, in the search for the hydrogen storage materials, the introduction of *i.a.* Al, Si, Ti, V, Cr, Mn, Fe, and Ga in the place of Co was considered. For systems in which part of Co atoms was replaced by Al, Si, Cr, or Ti, the *ab initio* calculations were performed. The substitutional disorder was treated in the coherent potential approximation (CPA) [16,17] or with the supercell method, with structures containing eight formula units (f.u.) containing one or two substituted Co atoms [18,19]. Our previous work [20] presents both the experimental and theoretical preliminary results for the melt-spun samples of the pure and Ti or Nb substituted $YCo_2$. There is an evidence of magnetically ordered phase at $T$ lower than 25 K [20, 22], which was linked with the structural disorder.

In this paper the main goal is to show the changes in the magnetic properties with increasing disorder. We initially introduced the disorder by rapid quenching. Further significant changes on the nanostructural/microstructural scale were caused by severe plastic deformation. We believe that enhanced magnetic ordering can be driven mainly by increased density of the lattice defects, as reported for example in [21]. The presented *ab initio* calculations are focused on the explanation of the influence of topological disorder (implemented through the presence of vacancies) and chemical disorder (analysis of Ti- and Nb-doped $YCo_2$ systems and native substitutions of Co in the place of Y and vice versa). The first part of the theoretical considerations refers to the experimental results showed in the present paper, while the second part concerns the experimental data from our previous paper [20]. The $Y_{0.9}Nb_{0.1}Co_2$ and $Y_{0.9}Ti_{0.1}Co_2$ are recalculated with the supercell method, instead of the CPA [20]. The novelty are the results for $YCo_2$ systems with vacancies and with dopants at Co site. The considered set of point defected structures is finally employed to evaluate the site preference energies for Nb and Ti substitutions. Calculations of binding parameters are crucial for the understanding of formation of the magnetic ground state in the amorphous structure and other disordered structures.

## Experiment and calculations

High purity Y and Co (3N) elements were used to prepare the master alloy of the nominal composition $YCo_2$ by arc-melting in Ar atmosphere. The elements were remelted several times to ensure homogeneity. Subsequently, the ingots were rapidly quenched in a melt-spinning device on a Cu wheel rotating with the surface velocity of 40 $ms^{-1}$. Afterwards, the high pressure torsion (HPT) was performed between two flat anvils under quasi-hydrostatic pressure of 4 and 6 GPa at room temperature on samples of 10 mm in diameter. The alloys were processed for 0.5 revolution with the rotational speed of 0.2 revolution per minute. Structural data were obtained by the X-ray diffraction (XRD) with the use of PANalytical X'Pert diffractometer in Bragg–Brentano geometry with Cu Kα radiation ($\lambda = 1.54056$ Å) over the $2\Theta$ range of 20º –100º, at 0.017º step size. Differential scanning calorimetry (DSC) isochronal experiments were performed using a Netzsch DSC 404 apparatus at the heating rate $q$ of 20 K/min and for temperatures from 50 to 900°C. The continuous heating curves were measured twice for each sample in a continuous argon flow (100 ml/min). Temperature and field dependences of magnetization were measured by a vibrating sample magnetometer (VSM) option in the Quantum Design Physical Property Measurement System.

The site preference energies for the substitutions were evaluated on the basis of *ab initio* calculations carried out for all point defected structures. For *ab initio* calculations we used the full-potential linearized augmented-plane-wave method (FP-LAPW) as implemented in the WIEN2k code [23] within the generalized gradient approximation (GGA) for the exchange-correlation potential in the Perdew-Burke-Ernzerhof form (PBE) [24]. Relativistic effects were included with the second variational treatment of the spin-orbit coupling. The calculations were performed for a plane wave cut-off parameter $RK_{max} = 7$ and total energy convergence criterion $10^{-6}$ Ry. Several crystal structures were considered: the perfect bulk $YCo_2$, two systems with vacancies (v@Y, v@Co), and four systems with substituted atoms

(Ti@Y, Ti@Co, Nb@Y, Nb@Co). The symbol “@” indicates that one atom from the supercell (Y or Co) is replaced by dopant atom (Nb or Ti) or by vacancy (v). In order to construct the basic crystal structure of $YCo_2$ the experimental lattice parameter ($a$ = 7.217 Å) and Wyckoff positions were used. The $YCo_2$ compound crystallizes in the $MgCu_2$-type structure with *Fd*-3*m* space group (no. 227). The unit cell used for the calculations contained six atoms divided into two groups: two Y atoms at 8a position (1/8, 1/8, 1/8) and four Co atoms at 16d position (1/2, 1/2, 1/2).

To reproduce atomic disorder the supercell method was used. The supercell was obtained by transforming the face-centered cubic system of $MgCu_2$-type structure into the primitive cubic structure (symmetry group P1) with 24 inequivalent atoms. The supercell contained then 8 formula units (f.u.) of $YCo_2$. Finally, to create structures with point defects, the atom of Y or Co was replaced by the dopant (Ti or Nb) or removed. As the supercell consisted of 8 Y and 16 Co atoms a replacement of single atom led to a concentration of dopant equal to 1/8 or 1/16, for example Ti@Y = $Y_{7/8}Ti_{1/8}Co_2$ = $Y_{0.875}Ti_{0.125}Co_2$. A more realistic arrangement of atoms was obtained by minimization of forces acting on the atoms for all supercells containing point defects. The numbers of k-points in the irreducible wedges of Brillouin zones were set to 641 irreducible k-points (20 x 20 x 20 mesh) for the bulk $YCo_2$, 256 (8 x 8 x 8) for the systems with vacancies and 864 (12 x 12 x 12) for the doped systems. For each structure the non-magnetic and spin-polarized calculations were performed. All spin-polarized calculations were started with antiparallel initial magnetic moments on Y and Co sublattices. However the ferromagnetic solution was not forced, as the magnetic moments at each inequivalent atom converged separately. For visualization of the crystal structures the VESTA code was used [25].

The site preference energy $\delta E_X$ for the substitution of X (Ti or Nb) in $YCo_2$ can be formulated, on the basis of Wolf’s work [26], as:

$$\delta E_X = [E_{tot}(X@Co) + E_{tot}(Co)] - [E_{tot}(X@Y) + E_{tot}(Y)]. \quad (1)$$

The site preference energy $\delta E_X$ was calculated by comparison of total energies for two complex systems with different sites doped. In the first complex system, the impurity X substitutes Co, which leads to the system composed of X@Co and a free atom of Co. The second system contains an analogous configuration, but with impurity X substituting Y atom of $YCo_2$. A subtraction of the total energies of these two systems gives information about which site is more energetically stable if substituted by the impurity X. An alternative approach to calculate the preferred sites of impurities has been proposed by Legoas and Frota-Pessôa [27]. They have suggested, for the Laves phase $ZrFe_2$, that upon introduction of impurity, the dopant atom locates on the previously existing vacancy. On the basis of this assumption, the new equation for the site preference energy for $YCo_2$ is written as:

$$\delta E_X = [E_{tot}(X@Co) - E_{tot}(v@Co)] - [E_{tot}(X@Y) - E_{tot}(v@Y)]. \quad (2)$$

## Results and Discussion

This section is divided into two parts, from which the first one is devoted to the experimental results for $YCo_2$ in various structural states, while the second one covers the results of calculations for $YCo_2$ with point defects.

### 1. Experiment

In its initial rapidly quenched state, the $YCo_2$ compound has $MgCu_2$-type structure (schematic view in Fig. 1), as shown in Fig. 2 and reported in Ref. [20]. A small peak at about 42.5° ($YCo_2$ pattern in Fig. 2) can be assigned to the atomic packing similar to that in the crystal structure of the $YCo_3$ phase (rhombohedral, $R$-3$m$ space group) which can be also presented as alternatively stacked structures of $YCo_5$ ($CaCu_5$-type) and $YCo_2$ ($MgCu_2$-type) at the 1:2 ratio. It is commonly reported that different Y-Co phases (for example $YCo_3$) contain some Co-rich regions [28]. It has been shown also by us [20,22] that the low temperature glassy behavior is present in the rapidly quenched $YCo_2$ and $(Y,Gd)Co_2$, and can be connected with the quenched-in disorder and the chemical segregation of Co.

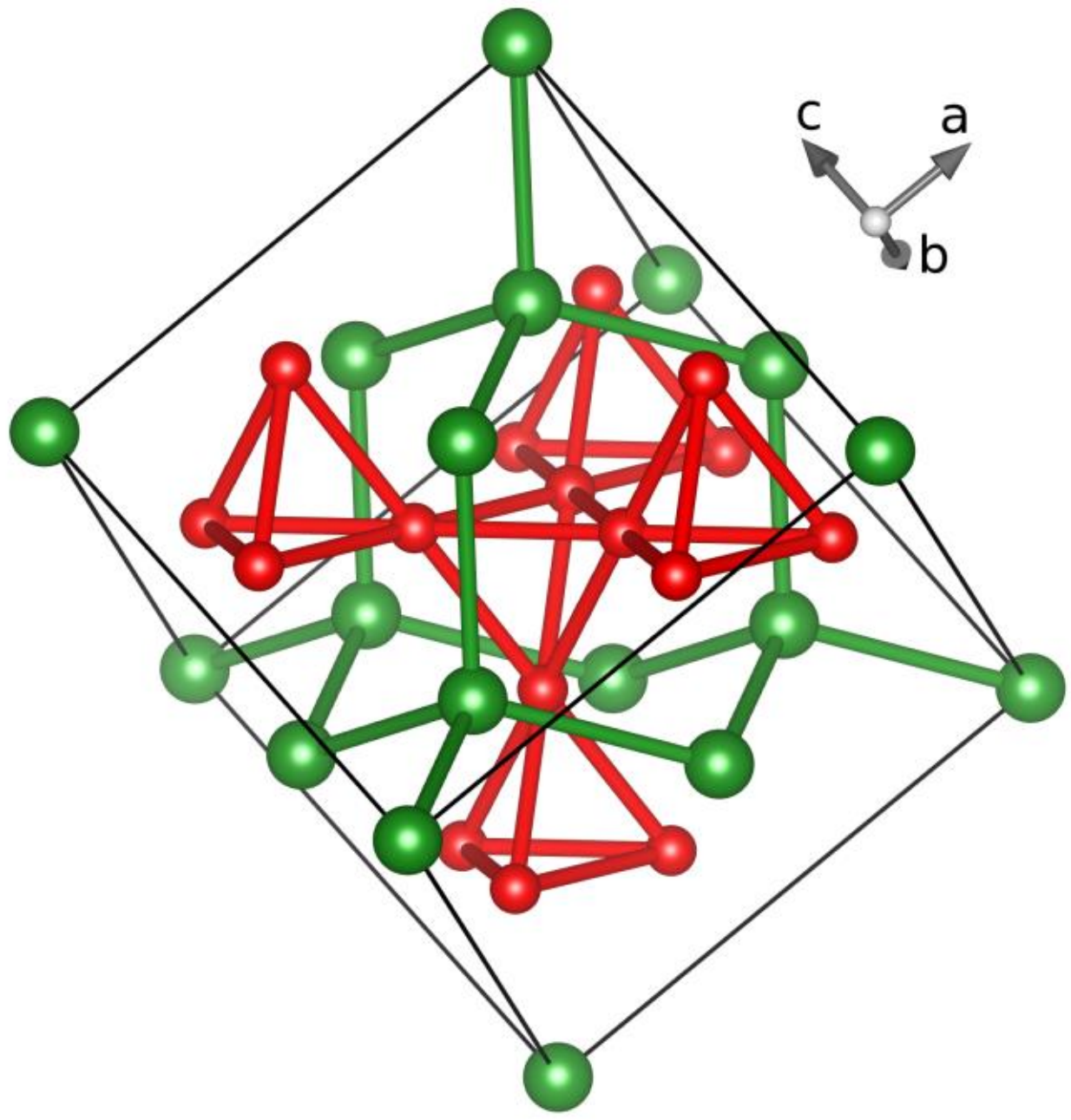

Fig. 1. Crystal structure of the cubic $MgCu_2$-type Laves phase with the large green balls representing Mg atoms and the small red balls indicating Cu atoms.

The X-ray diffraction patterns of the plastically deformed samples (using HPT at room temperature) indicated the presence of the main phase ($MgCu_2$-type structure), as shown in Fig. 2. A slight increase in the lattice parameter is observed for the samples after plastic deformation and it is suggested to be induced by lattice strain. Similar behavior has been already reported for irradiated $YCo_2$ thin films and ascribed exactly to the presence of stress [29]. Lattice constant *a* increases from 7.223 Å in the as-quenched state (in agreement with [13]) to 7.234 Å and 7.243 Å for the samples deformed at 4 and 6 GPa, respectively. The main difference between diffraction patterns is the broadening of diffraction peaks, which is usually connected with the increase in strain and refinement of crystallites [30]. Lattice strain was caused by the presence of crystal imperfections.

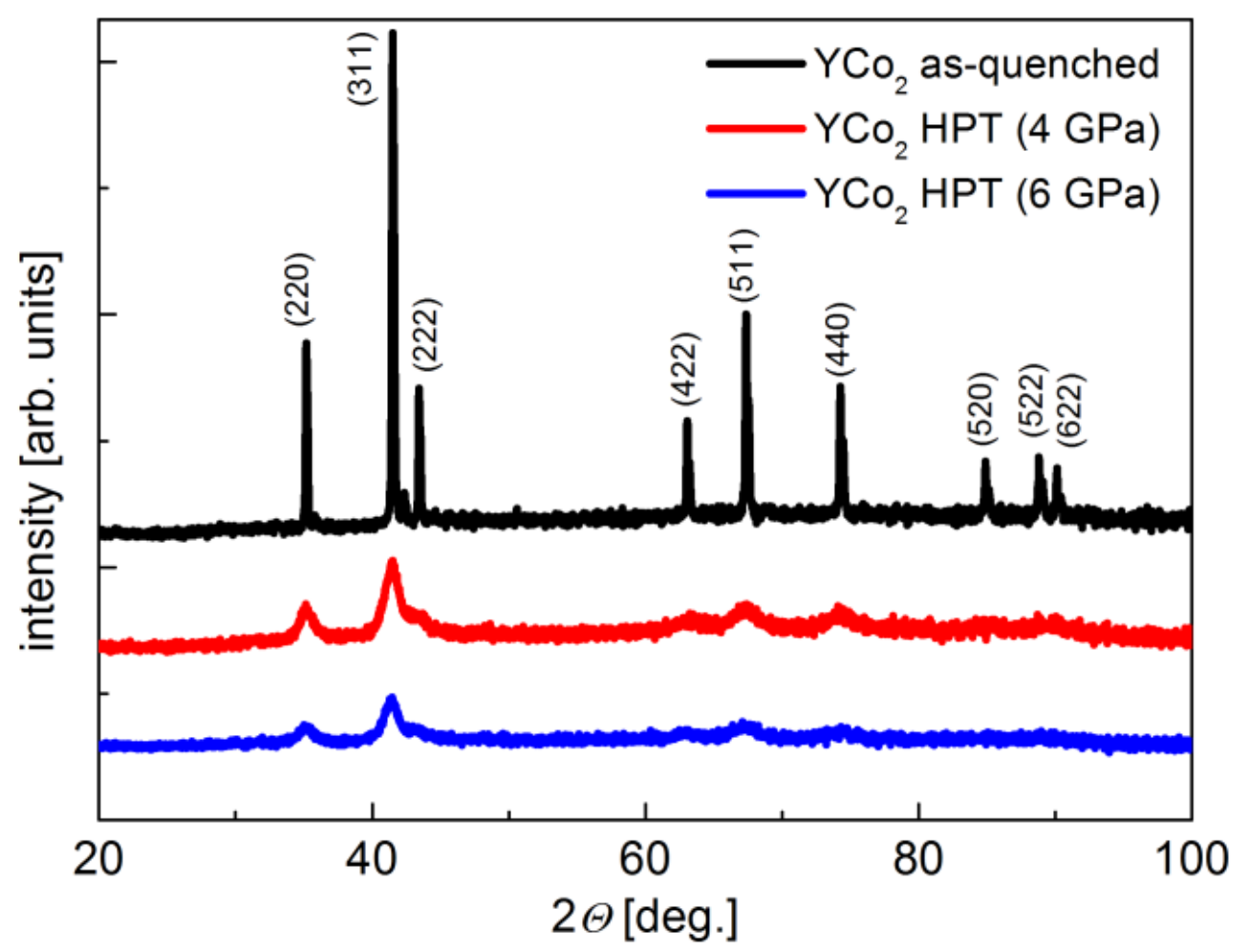

Fig. 2. X-ray diffraction patterns of $YCo_2$ ribbons in as-quenched state and after a plastic deformation (HPT - high pressure torsion) at 4 and 6 GPa.

We used the Williamson-Hall (W-H) method to extract the influence of the crystallite size, which changes as $1/\cos\Theta$, and strain, which varies as $\tan\Theta$ [31]. Full width at half maximum (FWHM) values were the basis for further calculations. The width of the Bragg peak is a combination of instrumental broadening and features characteristic of a particular sample. Therefore, it was necessary to exclude the instrumental contribution at first. Instrument-corrected broadening $\beta$ (instrument-corrected FWHM) was calculated using the relation [32]:

$$\beta = (\beta^2_{measured} - \beta^2_{instrumental})^{1/2}, \tag{3}$$

where $\beta_{instrumental}$ was the diffraction pattern of a standard material $LaB_6$ and it was subtracted from the measured value $\beta_{measured}$. The influence of $K_{\alpha 2}$ was also taken into account. All of the reflections up to $2\Theta = 80^o$ were included in the W-H method analysis. Crystallite size ($D \sim 1/(\beta_D \cos\Theta)$) and strain ($\varepsilon \sim \beta_s/\tan\Theta$) depend on $\Theta$ in different ways, as mentioned, and are the additive components in the W-H method:

$$\beta = \beta_D + \beta_S = K\lambda/D\cos\Theta + 4\varepsilon\tan\Theta, \tag{4}$$

where $K$ is the shape factor (assumed in this paper as $K = 0.9$) and $\lambda$ is the wavelength. Equation 4 can be rewritten in the form:

$\beta\cos\Theta = K\lambda/D + 4\varepsilon\sin\Theta.$ (5)

We extracted the strain and the crystallite size from the slope and intersection of the linear fit made to the plot of $4\varepsilon\sin\Theta$ dependence of $\beta\cos\Theta$ [33], respectively. The values of crystallite size and microstrain determined on the basis of W-H method are listed in Tab. 1.

Tab. 1. The estimated microstructural parameters of $YCo_2$ ribbons in as-quenched (as-q) state and after plastic deformation at 4 and 6 GPa.

| | $YCo_2$ (as-q) | $YCo_2$ (4 GPa) | $YCo_2$ (6 GPa) |
|---|---|---|---|
| crystallite size $D$ (nm) | 129 ± 20 | 32 ± 23 | 25 ± 21 |
| lattice strain $\varepsilon \times 10^{-3}$ | 0.15 ± 0.06 | 11.51 ± 2.12 | 11.84 ± 2.13 |

After the HPT processing (at 4 and 6 GPa) the mean size of the crystallites decreased significantly. Such behavior has been reported also for other systems [34–36]. We cannot exclude the formation of amorphous structure (with a magnetic moment of about 1 $\mu_B$/Co atom as reported in [5]) on the basis of X-ray diffraction measurements. However, in the differential scanning calorimetry (DSC) experiments (Fig. 3) no evident and abrupt changes were observed. Moreover, the first and second continuous heating curves for all samples were almost identical, suggesting the absence of any irreversible transitions, as for example crystallization. Therefore, the presence of amorphous regions is not discussed in the paper. Nevertheless, sensitivity of DSC could be insufficient to detect small volume fractions. XRD analysis suggests an increase in the lattice strain due to the deformation process up to $\varepsilon$ above $11\times10^{-3}$, which is consistent with the alteration in the lattice constant. The crystallite size of the as-quenched sample (exceeding 100 nm) was on the verge of applicability of the W-H method, due to the fact that the instrumental peak width (FWHM = 0.07º) was comparable to the crystallite size broadening. Therefore, changes in the crystallite size and lattice strain should be taken into account rather qualitatively, than quantitatively, and our intention was to show the influence of HPT and to confirm its expected impact on both parameters.

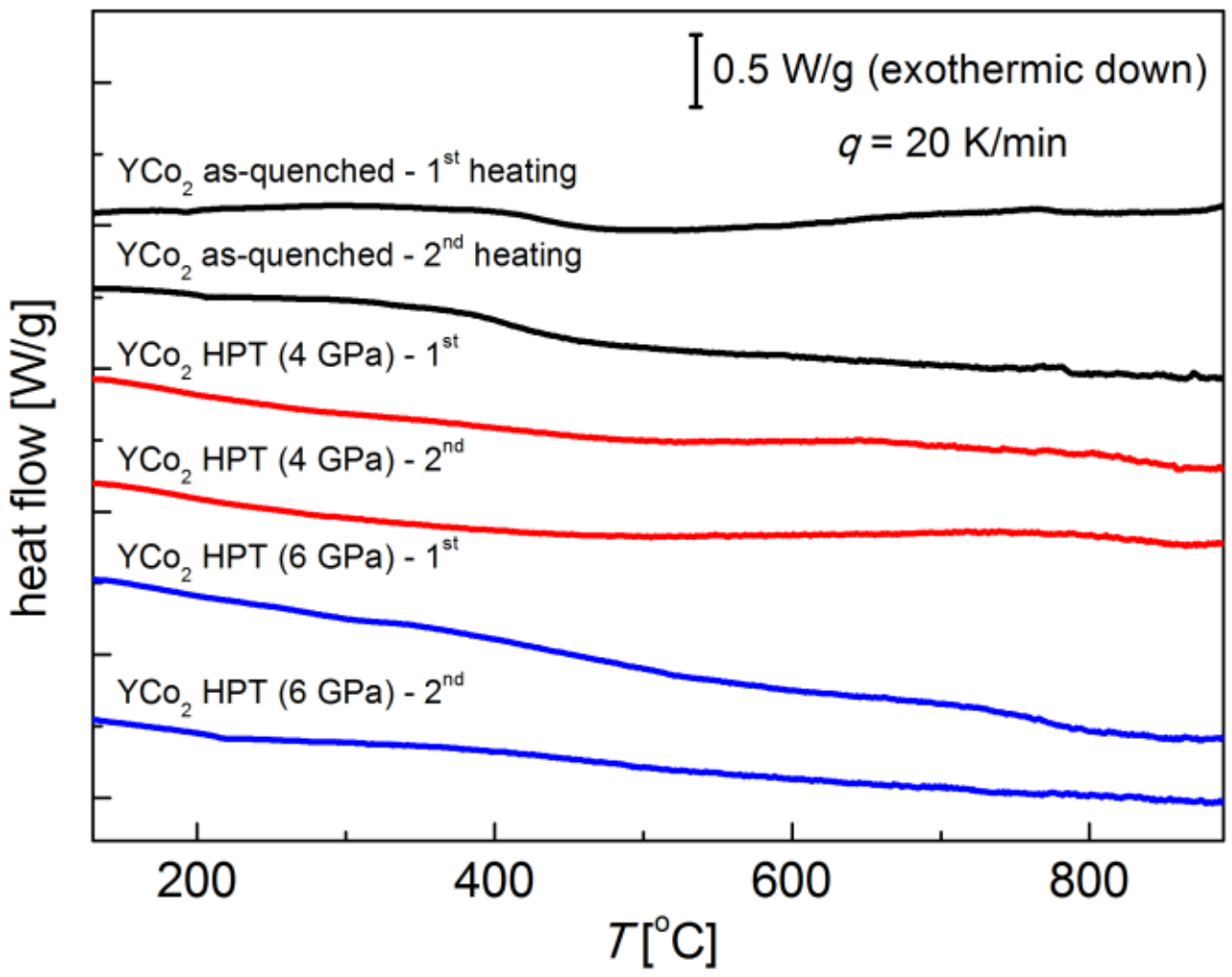

Fig. 3. The continuous heating differential scanning calorimetry curves of $YCo_2$ ribbons in as-quenched form and after plastic deformation (HPT – high pressure torsion) at 4 and 6 GPa measured at the heating rate of 20 K/min.

The reduced coordination at the surfaces, interfaces and grain boundaries led to the phenomenon of surface induced magnetism in $YCo_2$ [37]. Room temperature soft ferromagnetism has been also reported for nanocrystalline ball-milled $YCo_2$, as arising from the core of nanocrystals [13]. The ferrimagnetism of amorphous $YCo_2$ has been connected with a smaller density and Co-richer surroundings of Co atoms, when compared to the crystalline counterpart. The increase in the volume fraction of grain boundaries and occurrence of amorphous regions should significantly influence the magnetic properties of the deformed samples. We have shown that structural disorder (topological and chemical) can also deteriorate Pauli paramagnetism by formation of regions with spin-glass-type behavior [20,22]. In this paper we consider the as-spun sample with quenched-in disorder. Slight changes are visible even in the $M(T)$ curve (Fig. 4), as the low-temperature feature that appeared aside from the broad maximum typical of the exchange-enhanced paramagnets [37,38]. X-ray diffraction measurements imply that we should observe clear signs of the induced magnetic ordering due to the presence of refined and structurally disordered grains after plastic deformation. Magnetic measurements were performed to confirm this assumption and to show the magnitude of this transformation.

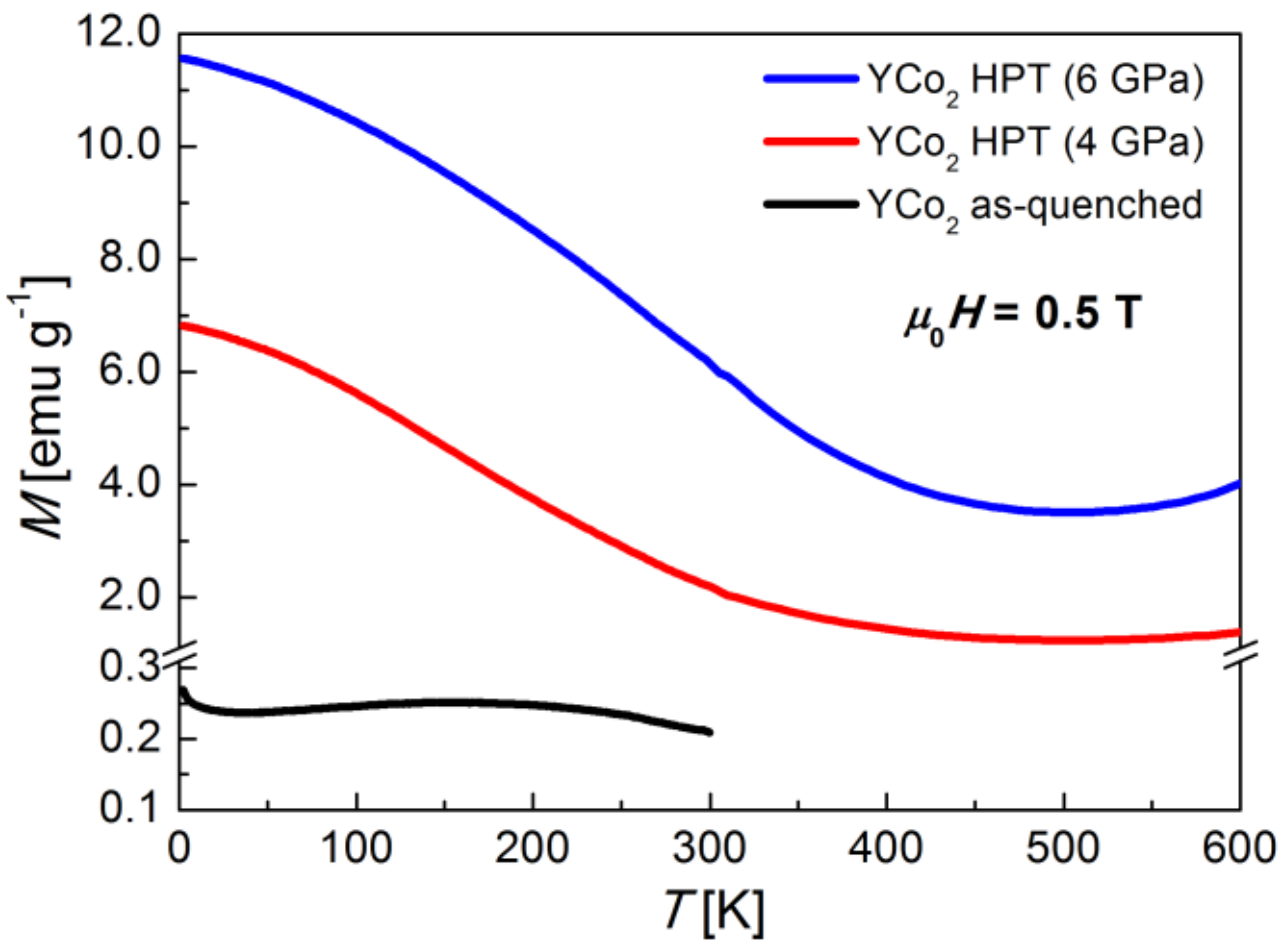

Fig. 4. Temperature dependence of the magnetization of $YCo_2$ ribbons in the as-quenched form and after plastic deformation (HPT – high pressure torsion) at 4 and 6 GPa.

The thermal variation of magnetization of $YCo_2$ ribbons in as-quenched form and after high pressure torsion at 4 and 6 GPa is plotted in Fig. 4. We recorded the curves during heating after field cooling in a magnetic field of 0.5 T. This value was chosen to compare the present results with literature values [13,20,39]. As mentioned, the presence of maximum in the $M(T)$ curve measured in constant field is typical of exchange-enhanced paramagnets and was observed for the rapidly quenched sample. In turn, the low temperature upturn is unique and has already been analysed [22]. A related maximum at about 10 K was observed in the temperature dependence of the AC magnetic susceptibility. This maximum was frequency dependent, which confirmed the occurrence of the spin glass features. There is a co-existence of regions with long and short range magnetic order in the rapidly quenched sample, as the real part of AC susceptibility did not vanish for temperatures up to 340 K. Both effects were connected with the presence of structural disorder [20,22]. Nevertheless, its impact was very limited and barely noticeable. Bearing in mind that the $M(T)$ curve for the as-quenched $YCo_2$ is presented in the enlarged scale, the magnetization values at 2 K for plastically deformed samples are more than one order of magnitude higher. Temperature dependences of magnetization of both samples after high pressure torsion are characteristic of ferromagnetic materials. The magnetization values increase again above room temperature. It could be connected with the formation of other magnetic phases, but further analysis is out of the scope of this paper.

$YCo_2$ consists of two magnetic sublattices, where the Co sublattice has an intrinsic magnetic moment, while the moment on Y is induced and has the opposite sign. We call such magnetically ordered state ferrimagnetic (FiM), while magnetic state of $YCo_2$ with zero moment is paramagnetic (PM) when considering experiment and non-magnetic (NM) when considering theoretical results. According to that, both plastically deformed $YCo_2$ samples are ferrimagnetic with some signs of paramagnetic/superparamagnetic behavior, as can be seen in Fig. 5. All samples are far from being saturated. In the bulk crystalline $YCo_2$ the saturation state could not be reached even above the metamagnetic phase transition [10]. All experimental values of magnetic moment per Co atom (Fig. 5) were calculated assuming non-magnetic Y atoms. The temperature and field dependences of magnetization are analogous to that of ball-milled nanocrystalline forms of $YCo_2$ [13]. In the polycrystalline $YCo_2$ the value of magnetization at the metamagnetic transition field has been reported to be equal to 0.17 $\mu_B$/Co and 0.44 $\mu_B$/Co in paramagnetic and ferrimagnetic state, respectively [10]. Moreover, $YCo_2$ reaches the highest value of magnetic moment of 1.0 $\mu_B$/Co in the amorphous form [5]. The fine particles obtained after ball-milling have the saturation moment (linear extrapolation of high field data to zero field) at 35 K equal to about 0.9 $\mu_B$/f.u. (~0.45 $\mu_B$/Co), so comparable to the magnetic moment in the polycrystalline bulk sample in the ferrimagnetic state. The magnetic moment values determined for our plastically deformed ribbons from $M(\mu_0 H)$ data at 2 T, reached 0.10 $\mu_B$/Co (sample deformed at 4 GPa) and 0.25 $\mu_B$/Co (6 GPa) at 2 K. The magnetic moments are of the same order of magnitude as those of literature examples and therefore they must arise not only from the surface of refined nanocrystals but also from the volume. The surface contribution to the magnetic moment in $YCo_2$ nanoparticles and in similar systems was estimated to be by a few orders of magnitude smaller [13,39,40]. The contributions from other magnetic phases, with free Co and oxides among them, are highly improbable, as we have no evidence of their presence and the results are consistent with those of the single phase $YCo_2$ (as cited in this paragraph).

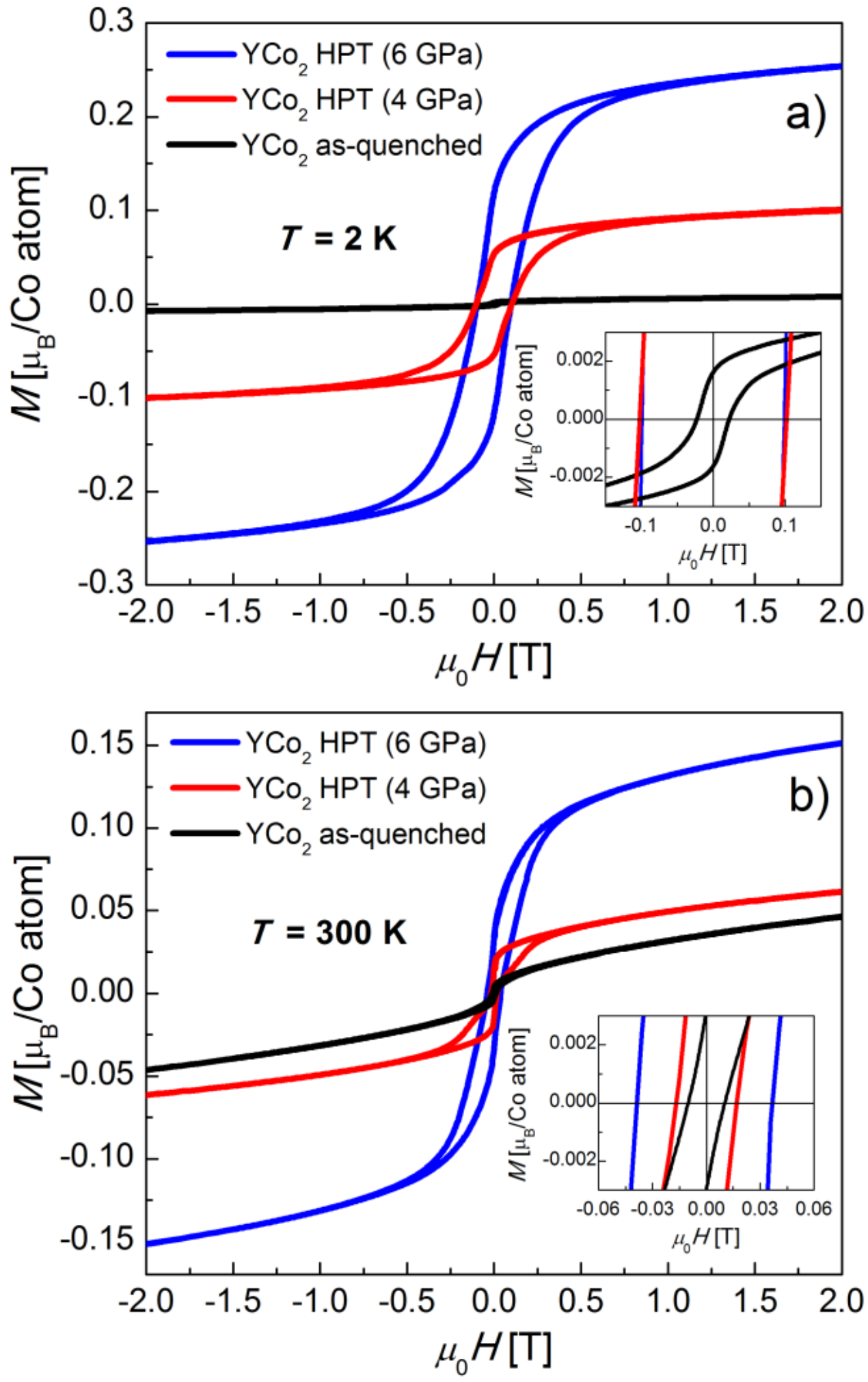

Fig. 5. Magnetic field dependences of magnetization of $YCo_2$ ribbons in as-quenched form and after plastic deformation (HPT – high pressure torsion) at 4 and 6 GPa measured at (a) 2 K and (b) 300 K.

The *M*/*H* value (measured in field cooled regime) of $YCo_2$ nanoparticles [13] has been reported to reach around 1 emu/mol at 2 K. In the present studies it reached half of this value for the sample deformed at 6 GPa, which is consistent with the *M*/*H* reported for example for ball-milled $LuCo_2$ nanoparticles [39]. The value of magnetic moment per Co atom increases with increasing severity of plastic deformation and is connected with the presence of nanocrystalline and/or disordered $YCo_2$ phase. Bulk $YCo_2$ has been reported to be paramagnetic, except for the two magnetically stable Co surface layers [37,38]. In turn, our nanoparticles were magnetic in their whole volume. We connect the stabilization of Co moment with the presence of structural disorder, namely the chemical disorder at crystallographic sites and other topological defects in crystalline structure. The increase in the lattice constant could be also responsible for the formation of long range magnetic order. All mentioned features can be responsible for the changes in Co-Co distances and for the fulfilment of the Stoner criterion as a result.

## 2. Density Functional Theory Calculations

*Ab initio* calculations were performed to make a comprehensive analysis of possible mechanisms of Co moment stabilization. Changes in lattice constant and the presence of vacancies can be compared with the present experimental results, while the Nb and Ti substituted compounds were investigated to explain the experimental results reported in our previous work [20]. Furthermore, in order to evaluate the site preference energies of point defects in $YCo_2$ we calculated the total energies of the substituted $YCo_2$ systems: Ti@Y, Ti@Co, Nb@Y, Nb@Co and for the $YCo_2$ with vacancies: v@Y and v@Co.

### *2.1. Properties of bulk $YCo_2$*

To accurately reproduce the band structure of bulk $YCo_2$ we started with the PBE calculations of the magnetic ground state and equilibrium lattice parameter. Figure 6 indicates that the $YCo_2$ magnetic ground state is magnetically ordered with the energy difference $\Delta E_{tot}$ between the equilibrium NM and FiM states equal to about 0.07 eV/f.u., see Tab. 2. The calculated equilibrium lattice parameters are $a_{NM}$ = 7.161 Å and $a_{FiM}$ = 7.227 Å, in comparison to the experimental value of 7.217 Å for the sample in paramagnetic state [9]. The resultant FiM ground state is in contradiction with the experimental paramagnetic ground state mainly because of the oversimplifications of the NM model, which only reproduces zero total magnetic moment and disregards non-zero local moments.

Tab. 2. The total energy difference $\Delta E_{tot}$ (eV/f.u.) between NM and FiM solutions for the $YCo_2$–based systems. A positive value indicates that magnetically ordered state is preferred. The systems designations are explained in the text.

| system | $YCo_2$ | v@Y | v@Co | Ti@Y | Ti@Co | Nb@Y | Nb@Co |
|---|---|---|---|---|---|---|---|
| $\Delta E_{tot}$ | 0.07 | 0.10 | 0.03 | 0.23 | 0.01 | 0.97 | 0.14 |

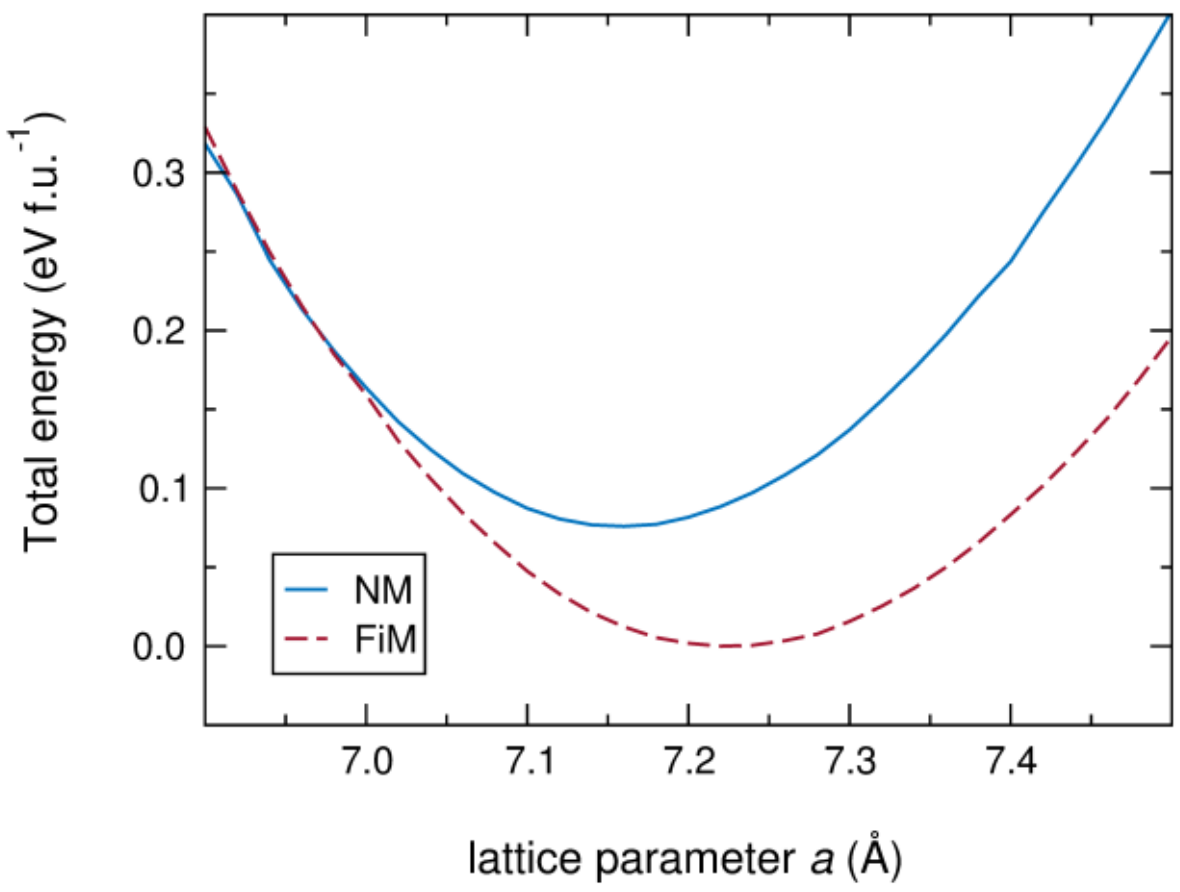

Fig. 6. The total energies calculated with FP-LAPW-PBE method *versus* lattice parameter. Results for FiM and NM solutions of the bulk $YCo_2$ are presented.

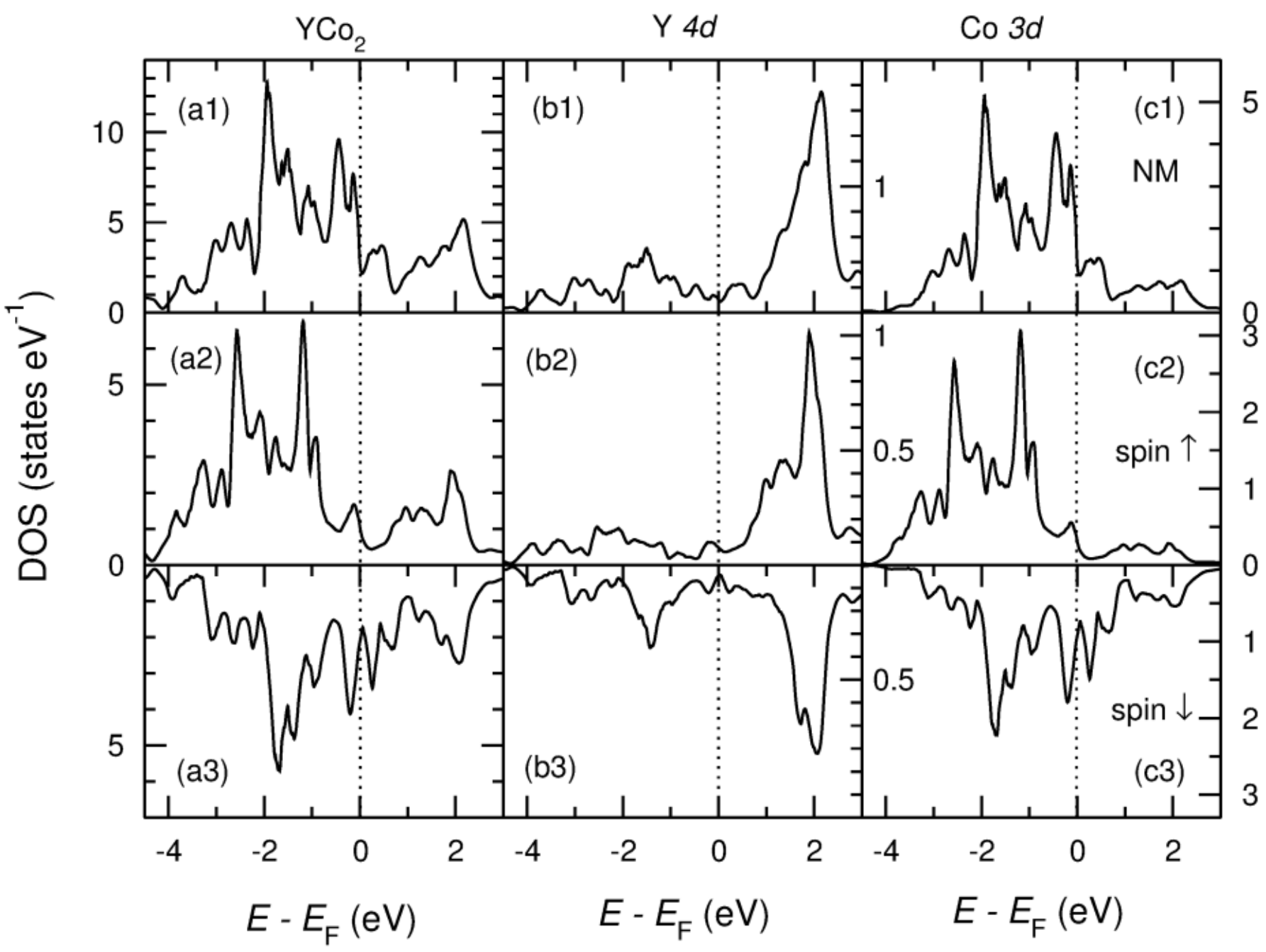

Fig. 7. Total and spin projected densities of states (DOS) for bulk $YCo_2$ phase calculated within the PBE approximation. Rows: (1) NM state; (2) FiM state – majority spin; (3) FiM state – minority spin. Columns: (a) total DOS; (b) Y 4*d* orbitals; (c) Co 3*d* orbitals.

It is well known that in $YCo_2$ the Co 3*d* and Y 4*d* valence bands overlap and hybridize [11, 41]. The Co 3*d* band is half-filled and has a high peak in density of states (DOS) just below the Fermi energy level ($E_F$). The almost empty Y 4*d* band has, in contrast, a wide maximum in DOS above the $E_F$. The $E_F$ is located in a dip of DOS between double Co 3*d* peak. The DOS of bulk $YCo_2$ presented in Fig. 7 serves as a reference for the following results for the modified structures. A comparison of the NM and FiM DOS reveals the mechanism of the transition from non-magnetic into magnetically ordered state. In the case of bulk $YCo_2$ the value of total DOS at $E_F$, contributing to the Stoner criterion, sensitively depends on the $E_F$ position [42]. For the NM state a double peak belonging to Co 3*d* states is located just below $E_F$ and DOS($E_F$) has a low value of 2.22 states/(eV·f.u.), but even a small change in the structure (chemical or topological) may result in shift of $E_F$ towards a high slope of DOS and trigger the transition towards magnetically ordered state. In the FiM state, Y and Co atoms carry spin magnetic moments of -0.19 and 1.19 $\mu_B$, respectively. However, in the FP-LAPW method the on-site spin magnetic moments are calculated within atomic spheres and the additional interstitial spin magnetic moment (-0.30 $\mu_B$) needs to be added to get the total magnetic moment, which is equal to 1.60 $\mu_B$/f.u. For comparison, the magnetic moments calculated by Schwarz and Mohn are -0.28 $\mu_B$ for Y and 1.02 $\mu_B$ for Co, in $YCo_2$ [43]. It is predicted from experiments that at full saturation the magnetic moment on Co will obtain value of about 1.0 $\mu_B$ [10].

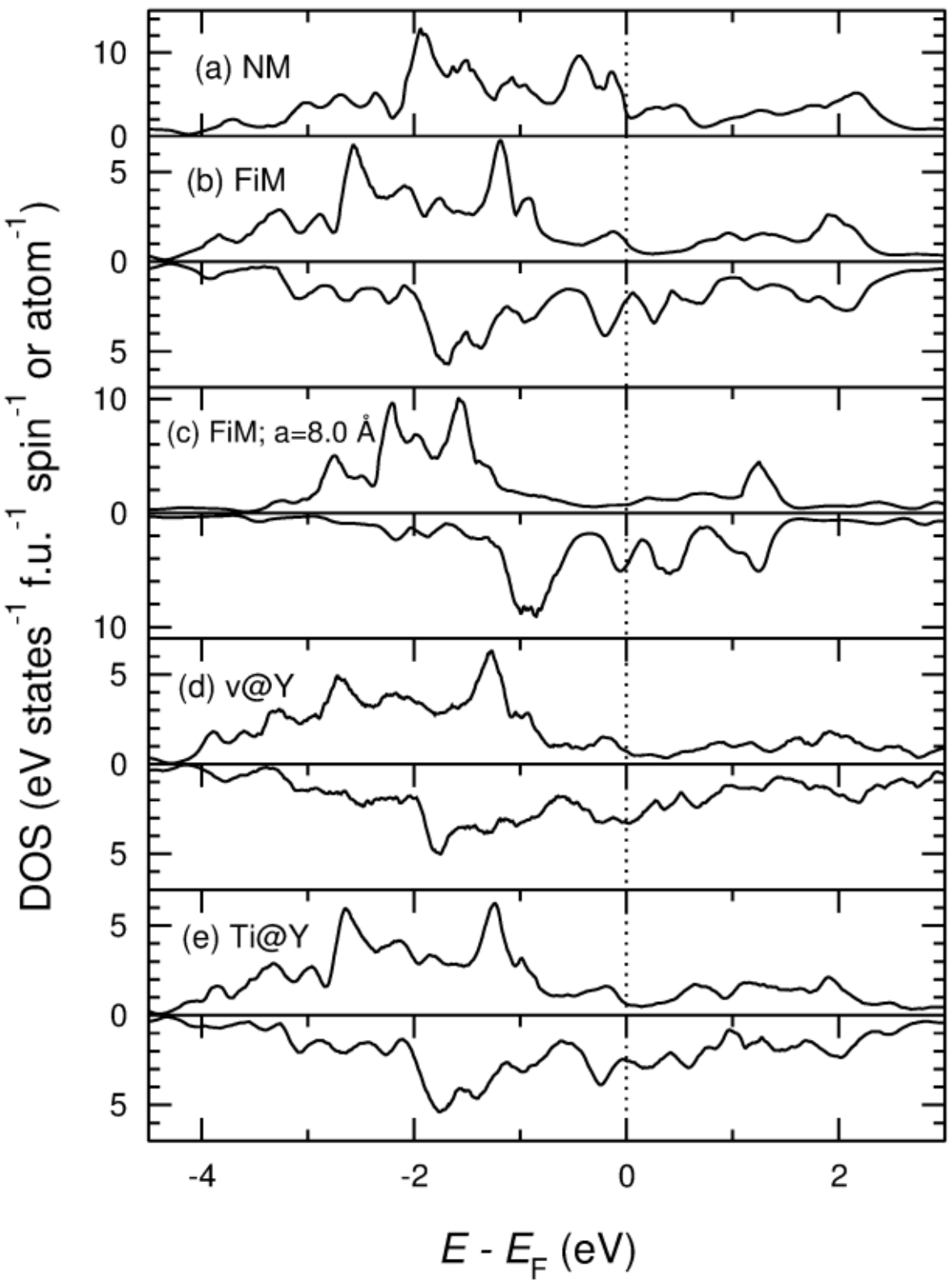

Fig. 8. Total and spin-projected DOS for $YCo_2$. (a) NM state, (b) FiM state, (c) FiM state with $a$ = 8.0 Å, (d) v@Y – FiM state, (e) Ti@Y – FiM state. Systems designations are explained in the text.

The growing tendency towards magnetism with increasing lattice parameter is a known behavior of $YCo_2$ [38]. Figure 8(c) presents a hypothetical FiM solution for the bulk $YCo_2$ phase with the lattice parameter $a$ = 8.0 Å. The observed narrowing of electronic bands is a result of localization of the valence electrons. For Y-Co phases, a narrowing of the Co 3*d*- bands is also observed with increasing Co concentration [3,43,44]. For FiM system with $a$ = 8.0 Å the magnetic moment $m_Y$ decreases from -0.19 to -0.32 $\mu_B$ and $m_{Co}$ increases from 1.19 to 1.50 $\mu_B$, which leads to the total magnetic moment of 1.95 $\mu_B$/f.u. Also the increase in Co content leads to the transition into the magnetic ground state. Our further results suggest the possibility of stabilization of the magnetically ordered state by suitable substitutions, which would enhance the lattice parameter without increasing Co concentration.

*2.2. Vacancies at Y and at Co sites*

Total energy differences $\Delta E_{tot}$ between NM and FiM solutions for $YCo_2$ supercells with vacancies are listed in Tab. 2. $\Delta E_{tot}$ is equal to 0.10 eV/f.u. for v@Y, and 0.03 eV/f.u. for v@Co system, in comparison to $\Delta E_{tot}$ for bulk $YCo_2$ equal to 0.07 eV/f.u. It indicates that the removal of Y atom (v@Y) stabilizes the magnetic ground state. The observed tendency is in agreement with the experimental observations and confirms that magnetic transitions in $YCo_2$ can be driven by structural modifications.

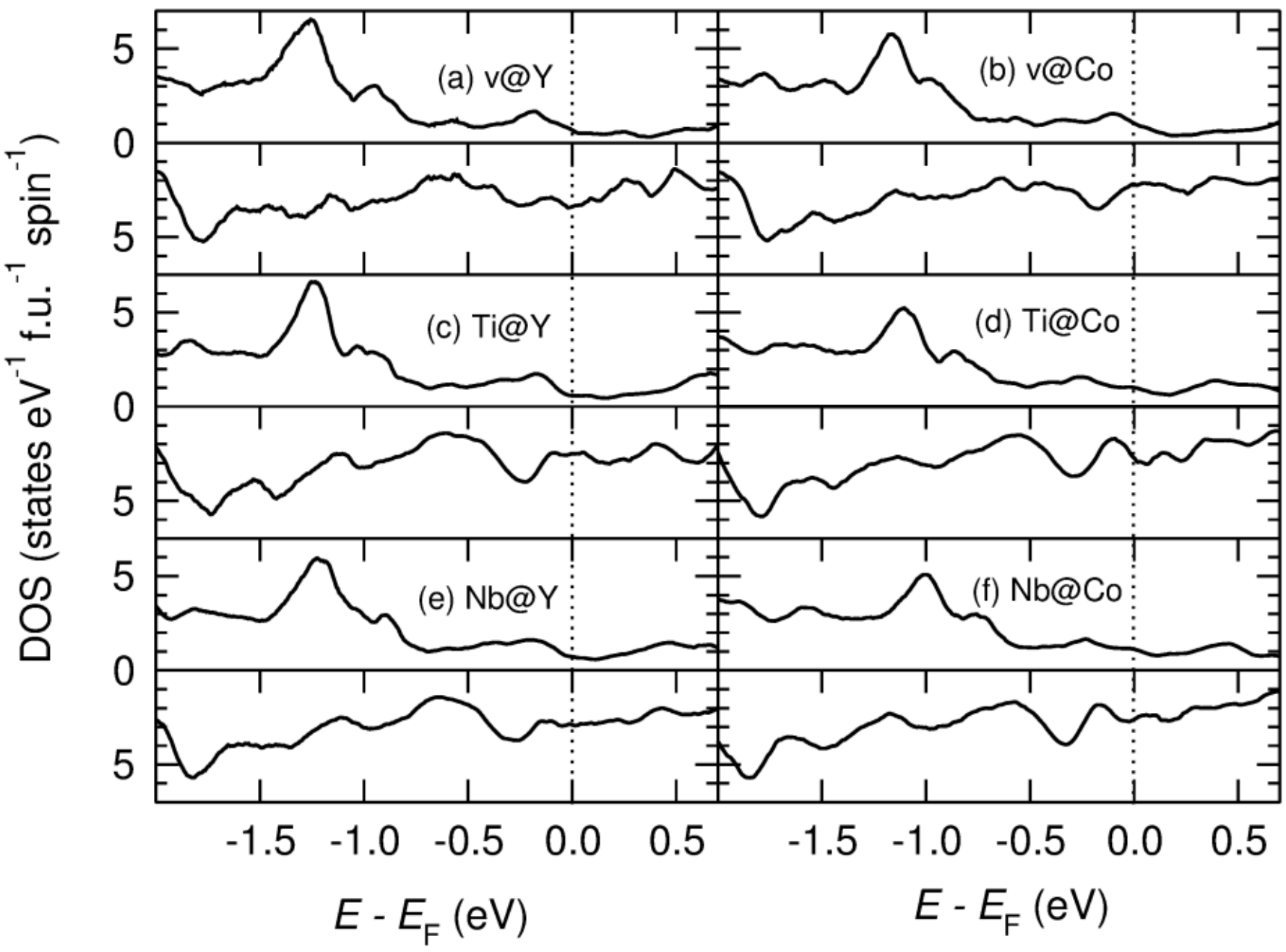

Fig. 9. Spin projected total DOS for $YCo_2$ systems with point defects. (a) v@Y, (b) v@Co, (c) Ti@Y, (d) Ti@Co, (e) Nb@Y, (f) Nb@Co. Systems designations are explained in the text.

The general features of DOS for systems with defects (Fig. 9) are similar to the solution for the bulk FiM. The main differences are observed in the vicinity of $E_F$.

Tab. 3. Spin magnetic moments $m$ ($\mu_B$) (within atomic spheres). Atoms are divided into types (*e.g.* Y1 and Y2) according to their values of $m$. The number of atoms of each type is listed. For bulk FiM $YCo_2$ $m_Y$ = -0.19 $\mu_B$ and $m_{Co}$ = 1.19 $\mu_B$.

| Atom | Y1 | Y2 | Co1 | Co2 | Co3 | Ti/Nb |
|---|---|---|---|---|---|---|
| Amount | 4 | 3 | 12 | 4 | | 0/1 |
| v@Y | -0.18 | -0.19 | 1.13 | 1.44 | | |
| Ti@Y | -0.17 | -0.18 | 1.07 | 1.11 | | -0.66 |
| Nb@Y | -0.15 | -0.17 | 1.00 | 1.12 | | -0.38 |
| Amount | 6 | 2 | 6 | 6 | 3 | 0/1 |
| v@Co | -0.17 | -0.16 | 1.05 | 1.01 | 1.04 | |
| Ti@Co | -0.13 | -0.14 | 0.74 | 0.92 | 1.02 | -0.20 |
| Nb@Co | -0.10 | -0.13 | 0.59 | 0.83 | 0.92 | -0.10 |

Spin magnetic moments $m$ for systems with defects are presented in Tab. 3. The positions of the two types of Co atoms in the v@Y supercell ($Y_7Co_{16}$) are shown in Fig. 10. The sixteen Co atoms in the unit cell form five tetrahedra with Co at their vertices. The Co-Co distances in one tetrahedron are reduced from 2.51 to 2.44 Å. The magnetic moment of Co2 atoms with smaller Co-Co distances increases from about 1.19 to 1.44 $\mu_B$. The relaxation of Co2 positions is schematically shown in Fig. 11(a).

In the v@Co system, see Tab. 3 and Fig. 11(b), the changes in the crystallographic neighborhood are much smaller than those in the v@Y system. The positions of the Co2 atoms do not change much because they are bound to an intact rigid sublattice of Y. The most significant impact on the properties of the v@Co system comes from a reduction in the number of Co-Co nearest neighbors from 6 to 5 for Co2 atoms surrounding the vacancy.

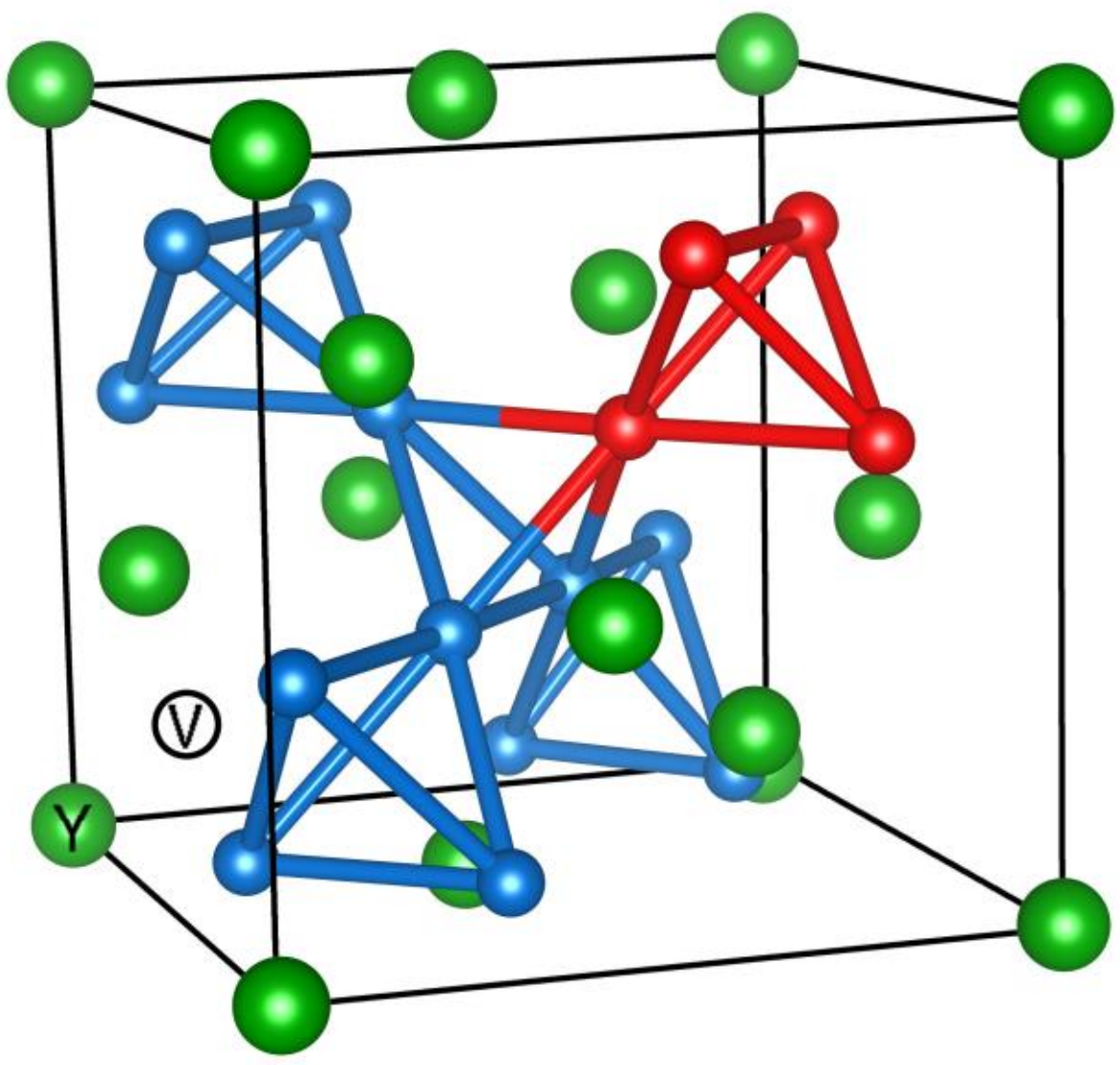

Fig. 10. The supercell of $YCo_2$ with a vacancy in the place of Y atom v@Y ($Y_7Co_{16}$). The Co atoms are separated into two types with the spin magnetic moments $m_{Co}$ = 1.13 $\mu_B$ (blue) and 1.44 $\mu_B$ (red).

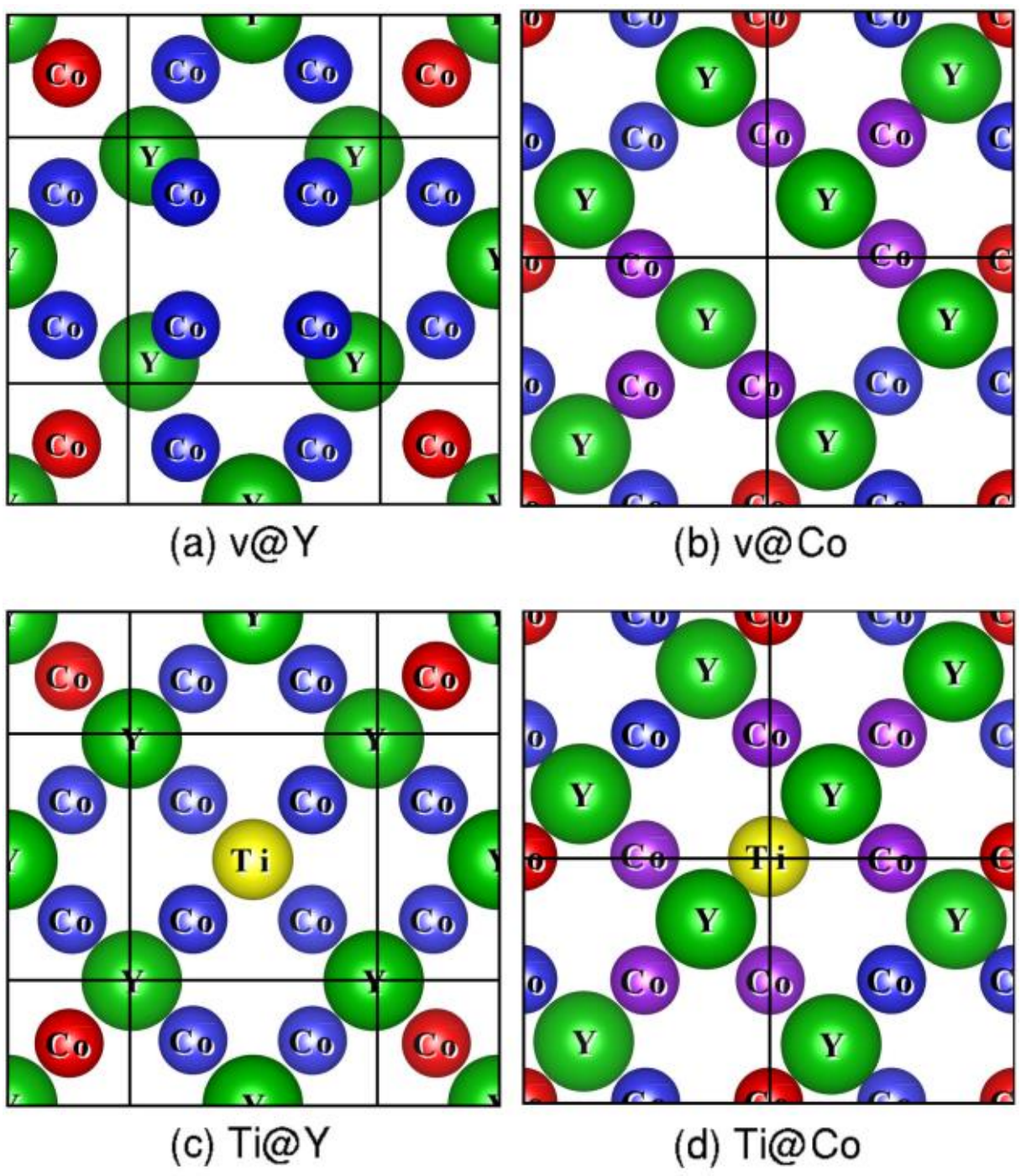

Fig. 11. Face views of $YCo_2$ structures with point defects: (a) v@Y; (b) v@Co; (c) Ti@Y; (d) Ti@Co. The colors of Co atoms correspond to different values of magnetic moments, see Tab. 3. Each representation is centered on a the specific defect. The considered structures with Nb substitutions do not differ noticeably from the Ti substituted ones, thus they are omitted in the graph.

*2.3. Doped structures: Ti@Y, Ti@Co, Nb@Y, Nb@Co*

Four structures with Y and Co atoms substituted by Ti or Nb are modeled. The notation used *e.g.* for the doped $YCo_2$ supercell with Y atom substituted by Ti is: Ti@Y or $Y_7TiCo_{16}$ or $Y_{0.875}Ti_{0.125}Co_2$. After substitution, the geometry is optimized, but the relaxed atomic positions do not change much, in contrast to those in the previously considered systems with vacancies. Thus, the electronic structure is affected mostly by the change in chemical composition.

The total energy differences $\Delta E_{tot}$ between NM and FiM solutions for doped $YCo_2$ systems are shown in Tab. 2. FiM is the favorable ground state for all four cases, however, higher stabilization of the magnetic state is obtained by the substitution of Y atoms. The predicted FiM ground states are in agreement with the experimental results for $Y_{0.9}Ti_{0.1}Co_2$ and $Y_{0.9}Nb_{0.1}Co_2$ [20].

The valence electrons configurations of atoms constituting the doped systems are: $4d^15s^2$ for Y, $3d^74s^2$ for Co, $3d^24s^2$ for Ti, and $4d^45s^1$ for Nb. The replacement of Co atom by Ti or Nb dopant is equivalent to removal of 5 or 4 electrons from the valence band. It lowers the bands filling and shifts $E_F$ towards lower energies. The replacement of Y by Ti or Nb dopant is like adding 1 or 2 electrons to the valence bands. It increases the bands filling and shifts $E_F$ towards higher energies, see Fig. 9.

For all doped systems the dopant atoms carry negative magnetic moments (see Tab. 3). For the systems with substituted Y atom, the moment on the dopant is more negative than that on Y. The Ti@Co notation means the $YCo_{1.875}Ti_{0.125}$ composition. It was found experimentally that for the $YCo_{1.9}Ti_{0.1}$ system the effective moment $m_{eff}$ is reduced from 3.86 to 3.4 $\mu_B$/atom, in comparison to that of pure $YCo_2$ [18].

*2.4. Binding properties of doped structures*

Site preference energies for the Ti or Nb dopants in $YCo_2$ and for the native point defects Y and Co are calculated using the Wolf and Legoas-Frota-Pessôa formulas. The results are shown in Tab. 4. All structures are calculated with spin polarization and geometry optimization. The positive values indicate the Y site preference. It should be emphasized that site preference energies calculated on the basis of the supercells method depend on the dopants concentration. The employed supercells contain 8 formula units with one substituted atom X and reproduce the compositions $Y_{0.875}X_{0.125}Co_2$ or $YCo_{1.875}X_{0.125}$.

Tab. 4. Site preference energies (in meV) for $YCo_2$ system doped by Ti and Nb and by native substitutions Y and Co. Positive values indicate Y site preference, negative Co site preference.

| Model | Ti | Nb | Y | Co |
|---|---|---|---|---|
| Wolf [26] | 44 | 492 | 93 | -43 |
| Legoas-Frota-Pessôa [27] | 14 | 462 | 32 | -103 |

Site preference energies calculated on the basis of the Legoas-Frota-Pessôa approach are in each case lower than those based on the Wolf's model. It means that for Ti and Nb impurities it would be easier to replace the occupied Y site than to fill Y vacancy. Site preference energies are comparable to, or higher than, the thermal energies at temperatures $T > 500$ K, at which bulk diffusion is still efficient. The Co sublattice will not be directly affected by the presence of Ti and Nb alloying elements, and the effects on the magnetic ordering will be dictated by the steric effect and ensuing lattice relaxation combined with modifications of the spin-split band structure in this itinerant magnet.

## Conclusions

Ferrimagnetic ordering was induced in bulk $YCo_2$ by severe plastic deformation, leading to a formation of bulk (non-powder) nanocrystalline alloys. Ferrimagnetic state was stable in the broad temperature range, even well above the room temperature. Broadly defined structural disorder (presence of grain boundaries, vacancies, chemical disorder) is believed to be the origin of induced Co magnetic moment, as confirmed also on the basis of *ab initio* calculations. The present bulk form of the nanocrystalline $YCo_2$ is superior to the nanocrystalline ball-milled samples reported before as it permitted elimination of the problems with oxidation and the presence of surfactants. Systematic results of *ab initio* electronic structure calculations for the $YCo_2$ Laves phase compounds with point defects are presented in the second part of the paper. The complete set of results for structures with vacancies and substitutions allows the calculation of site preference energies. Evaluation by Wolf and Legoas-Frota-Pessôa equations predicted that Ti and Nb impurities preferred to occupy the Y site. All considered point defects in place of Y lead to energetically more stable FiM ground states than with point defects in place of Co. This trend is in agreement with the experimental data on $YCo_2$ samples with Y atoms substituted by Ti and Nb. Another theoretical result confirmed experimentally is that Ti point defects in the place of Co lower the magnetic moment of the system. The calculated magnetic ground states are in qualitative agreement with experimental results for all considered point defect structures. The calculated magnetic moments at Co atoms were always close to 1.0 $\mu_B$ and opposite to smaller moments localized at Y and dopants. Enhancement of the lattice parameter also leads to an induced magnetic ordering.

**Acknowledgments**

The authors gratefully acknowledge the financial support of DAAD/MNiSW Project Based Personal Exchange Program (PPP) in years 2016–2017. M.W. and B.W. acknowledge the financial support from the Foundation of Polish Science grant HOMING. The HOMING programme is co-financed by the European Union under the European Regional Development

Fund. The authors would like to thank Ján Rusz, Erna Delczeg-Czirjak and Alexander Edström for helpful discussions and comments.